\documentclass[prl,12pt,floatfix]{revtex4}
\usepackage{graphicx}
\usepackage{amsmath}
\usepackage{amsmath}
\usepackage{amssymb}
\usepackage[mathcal]{euscript}

\begin{document}

\title{Towards understanding of birds magnetoreceptor mechanism}

\author{Ilia A. Solov'yov}
\altaffiliation{On leave from the A.F. Ioffe Institute, St. Petersburg, Russia.}
\email[Email address: ]{ilia@fias.uni-frankfurt.de}
\author{Walter Greiner}
\affiliation{Frankfurt Institute for Advanced Studies, Johann
Wolfgang Goethe University, Max-von-Laue Str. 1, 60438 Frankfurt am
Main, Germany}

\begin{abstract}
In the present letter we suggest a new theoretical model for a
quantitative description of the magnetoreception mechanism in birds.
The considered mechanism involves two types of iron minerals
(magnetite and maghemite) which were found in subcellular
compartments within sensory dendrites of the upper beak of several
bird species. The analysis of forces acting between the iron
particles shows that the orientation of the external geomagnetic
field can significantly change the probability of the
mechanosensitive ion channels opening and closing. The performed
theoretical analysis shows that the suggested magnetoreceptor system
might be a sensitive biological magnetometer providing an essential
part of the magnetic map for navigation.
\end{abstract}

\maketitle

A large variety of animals possess a magnetic sense. The
best-studied example is the use of the geomagnetic field by
migratory birds for orientation and navigation during migration.
Reviews of these studies are given in
Refs.~\cite{IliaGreiner07,SOLO2007}. In the present letter we
address this problem from the theoretical point of view. Based on
the known experimental observations we develop a physical model
which we use for the description of magnetoreception phenomena in
birds. The suggested model is based on the interaction of magnetic
particles consisting of ferrimagnetic iron-minerals magnetite
(Fe$_3$O$_4$) and maghemite ($\gamma-$Fe$_2$O$_3$), which were
observed in the beak of different bird species
\cite{Hanzlik00,Fleissner03,Fleissner07a}. Based on the analysis of
forces which act between these particles we show that the considered
iron-mineral system can serve as a magnetoreceptor with distinct
orientational properties. We demonstrate that- depending on the
orientation of the external magnetic field- the probability of
opening of mechanosensitive ion channels significantly changes, thus
leading to different nerve signals. The nerve signals are delivered
to the brain causing a certain orientational behavior of the bird.

\begin{figure}
\begin{center}
\includegraphics[scale=0.8,clip]{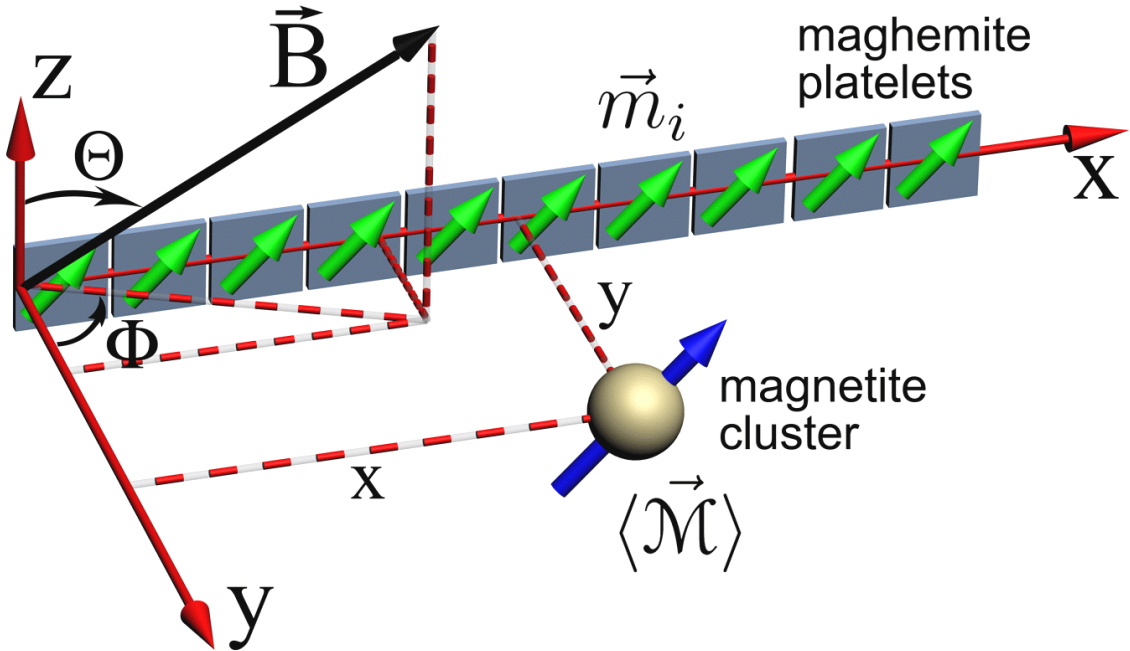}
\end{center}
\caption{Magnetoreceptor unit consisting of ten maghemite platelets
(boxes) and a magnetite cluster (sphere). The coordinate frame shown
here is used in the computations. The direction of the external
magnetic induction vector ${\vec B}$ is characterized by two polar
angles $\Phi$ and $\Theta$. The magnetic moments of the maghemite
platelet $i$, ${\vec m_i}$, and of the magnetite cluster $\langle
{\cal{\vec M}}\rangle$ are indicated} \label{system}
\end{figure}

The histology studies of the upper beak of homing pigeons
\cite{Fleissner03,Fleissner07a}  showed that iron minerals are
concentrated in six symmetrical spots near the lateral margin of the
skin of the upper beak inside the dendrites of nerve cells. For the
study of the magnetoreception function of the dendrite a primary
magnetoreceptor unit has been defined, being the smallest structure
possessing the magnetoreception properties of the whole dendrite.
The magnetoreceptor unit consists of ten maghemite platelets and one
magnetite cluster as shown in Fig.~\ref{system}. Experimental
observations \cite{Fleissner03,Fleissner07a} suggest that the
dendrite contains about 10-15 magnetoreceptor units, which should
have similar behavior in the external magnetic field. Therefore, if
the entire dendrite is subject to the external magnetic field the
repetition of the magnetoreceptory units increases the functional
safety of the dendrite magnetoreception.

\noindent

The geometry of the magnetoreceptor unit is determined from the
experimental observations \cite{Fleissner07a}. Thus, the maghemite
platelets have the dimensions $1\times0.1\times1$ $\mu$m and the
magnetite cluster has the diameter of 1 $\mu$m. The maghemite
platelets are located in the (xz)-plane being aligned along the
x-axis (see Fig.~\ref{system}). The distance between two neighboring
platelets is equal to 0.1 $\mu$m.

The size of a single maghemite platelet (MP) is sufficient for the
formation of magnetic domains in the (xz)-plane of the platelet
\cite{Kirschvink81} (see Fig.~\ref{system}). Thus, the MPs have a
magnetic moment in this plane even in the absence of the external
magnetic field. The magnetic moment of a platelet has the same
direction as that of the total magnetic field at its site, ${\vec
{\cal H}_i}$:

\begin{equation}
{\vec m_i} = Ml_xl_yl_z{\vec {\cal H}_i}/|{\vec {\cal H}_i}|,
\label{eq:maghemite_moment1}
\end{equation}

\noindent where $M$ is the remanent magnetization of maghemite,
$l_x$, $l_y$ and $l_z$ are the dimensions of a platelet along the
$x$, $y$ and $z$ axes respectively. With $M=50$ emu/cm$^3$
\cite{Mathe05}, $l_x=l_z=1$ $\mu$m and $l_y=0.1$ $\mu$m one obtains:
$m_i\approx3.121$ eV/G.

The magnetite cluster (MC) consists of nanoparticles which are 5 nm
in diameter \cite{Fleissner07a}. In the case of finite temperature
and finite magnetic field, the mean total moment of the MC,
$\langle{\cal{\vec M}}\rangle$, is:

\begin{equation}
\langle {\cal{\vec M}}\rangle\approx\frac{n\mu^2}{3kT}{\vec
H}=\chi{\vec H}, \label{eq:Tot_av_momentZ_approx}
\end{equation}

\noindent where $n$ is the number of nanomagnets in the cluster,
$\mu$ is the magnetic moment of an individual nanomagnet, ${\vec H}$
is the magnetic field strength at the site of the MC, $T$ is the
temperature and $k$ is the Bolzmann constant. With $R_0=0.5$ $\mu$m
and $r_0=2.5$ nm, being the radii of the MC and of the nanoparticle
respectively, one obtains $n\approx8\cdot10^6$.

The proportionality constant between the magnetic moment and the
field strength, $\chi$, is the magnetic susceptibility. Dividing it
by the volume of the MC one obtains the volume susceptibility of the
MC, $\chi_v$, which at $300$ $K$ is equal to 0.12 CGS units. With
$H=10$ Oe, which is the typical value of the total field at the site
of the MC, one obtains: $\langle {\cal M}\rangle\approx0.392$ eV/G.
Note that this value is about an order of magnitude smaller than the
magnetic moment of a single MP.

The potential energy of the MC reads as:

\begin{equation}
E({\vec R})=-\chi_v\frac{4}{3}\pi R_0^3\left|\frac{{\vec
B}}{\mu_{med}}+\sum_{j=1}^{N}{\vec H_{j}}({\vec R})\right|^2,
\label{eq:interaction_energy}
\end{equation}

\noindent where ${\vec R}$ defines the position of the MC, ${\vec
B}$ is the induction vector of the external magnetic field,
$\mu_{med}\approx 1$ is the permeability of the medium, $N$ is the
number of MPs, ${\vec H_{i}}({\vec R})$ is the magnetic field
created by the $i$-th MP at the site of the MC, which is known to be
\cite{IliaGreiner07}

\begin{equation}
{\vec H_j}({\vec R})=\frac{3\left({\vec R}-{\vec
r_j}\right)\left({\vec m_j}\left({\vec R}-{\vec
r_j}\right)\right)-{\vec m_j}\left|{\vec R}-{\vec
r_j}\right|^2}{\left|{\vec R}-{\vec r_j}\right|^5}.
\label{eq:maghemite_field}
\end{equation}

\noindent Here ${\vec r_j}$ describes the position of the j-th
platelet and ${\vec m_j}$ is its magnetic moment defined in
Eq.~(\ref{eq:maghemite_moment1}). The total magnetic field at the
site of the $i-$th MP is:

\begin{equation}
{\vec {\cal H}_i}=\frac{{\vec B}}{\mu_{med}}+\sum_{{j=1}\atop{j\ne
i}}^{N}{\vec H_{j}}({\vec r_i}). \label{eq:local_field}
\end{equation}

\noindent The first term describes the external magnetic field while
the second term describes the magnetic field created by all MPs
except the $i-$th one.

It follows from Eq.~(\ref{eq:local_field}) that the total magnetic
field ${\vec {\cal H}_i}$ is determined by the magnetic moments of
the platelets. Thus Eqs.~(\ref{eq:maghemite_moment1}) and
(\ref{eq:local_field}) have to be treated iteratively. In the
zeroth-order of approximation ${\vec m_i}$ are aligned along the
x-axis, what is energetically the most favorable configuration of
the system. The magnetic moment of a platelet is then $ {\vec
m_i^{(0)}}= Ml_xl_yl_z{\vec {\bf i}}$, where ${\vec {\bf i}}$ is the
unit vector along the x-axis. The total magnetic field in the
first-order approximation at the site of the $i$-th MP reads as:

\begin{equation}
{\vec {\cal H}_i}^{(1)}={\vec B}+2Ml_xl_yl_z\xi_i{\vec {\bf i}},
\label{eq:local_field_1order}
\end{equation}

\noindent where $x_i$ is the x-coordinate of the $i$-th platelet and
$\xi_i=\sum1/|x_i-x_j|^3$. Substituting
Eq.~(\ref{eq:local_field_1order}) into
Eq.~(\ref{eq:maghemite_moment1}) one yields the first-order
approximation for ${\vec m_i}$:

\begin{eqnarray}
m_{i_x}^{(1)}&=&\frac{Ml_xl_yl_z\left(B_x+2Ml_xl_yl_z\xi_i\right)}{\sqrt{\left(B_x+2Ml_xl_yl_z\xi_i\right)^2+B_y^2+B_z^2}}\\
m_{i_y}^{(1)}&=&0\\
m_{i_z}^{(1)}&=&\frac{Ml_xl_yl_zB_z}{\sqrt{\left(B_x+2Ml_xl_yl_z\xi_i\right)^2+B_y^2+B_z^2}}
. \label{eq:m_1order}
\end{eqnarray}

\noindent Here $B_x$ and $B_z$ are the x- and z-components of the
external magnetic induction vector respectively. In
\cite{IliaGreiner07} we demonstrated that the first-order
approximation can be used to calculate the interaction energy with
accuracy higher than 1\%.

\begin{figure}
\begin{center}
\includegraphics[scale=0.8,clip]{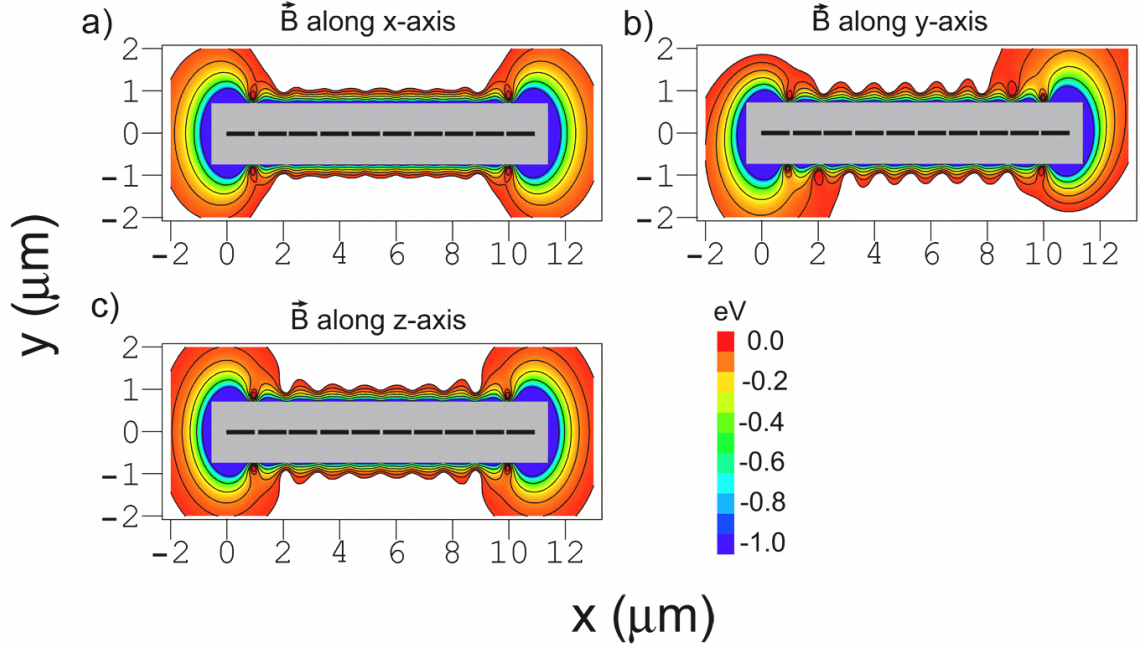}
\end{center}
\caption{Potential energy surfaces of the MC calculated as a
function of x and y coordinates, while z=0 $\mu$m (see
Fig.~\ref{system}) at different orientations of the external
magnetic field vector: plot a): magnetic field vector directed along
the x-axis; plot b): magnetic field vector directed along the
y-axis; plot c): magnetic field vector directed along the z-axis.
The MPs are shown with black rectangles. The gray rectangle in the
center of the potential energy surfaces shows the region, where the
MC can not be placed, due to its finite size.}
\label{fg:Energy_point}
\end{figure}

Figure \ref{fg:Energy_point} shows the potential energy surfaces of
the MC as a function of its coordinates x and y, while z=0 $\mu m$
(see Fig.~\ref{system}), calculated at different orientations of the
external magnetic field vector. Because of the MC size and because
of the MPs there exists a forbidden region on the potential energy
surface, where the MC can not be placed. The MPs are shown in
Fig.~\ref{fg:Energy_point} with black rectangles. The gray rectangle
in the center of the potential energy surfaces defines the forbidden
region for the MC.

\noindent

In our calculations the external magnetic field strength is 0.5 G,
being a typical value of the Earth magnetic field strength. The
potential energy surfaces (Fig.~\ref{fg:Energy_point}) were
calculated using Eq.~(\ref{eq:interaction_energy}). The potential
energy surfaces calculated for the external magnetic field directed
along the x-, y- and z-axes are shown in plots a), b) and c) of
Fig.~\ref{fg:Energy_point} respectively. The three potential energy
surfaces are similar, although some differences can be observed. The
potential energy surfaces corresponding to the x- and z-
orientations of the external field have axial symmetry along the
y=0, z=0 axis, while the potential energy surface corresponding to
the orientation of the external field along the y-axis has point
symmetry with respect to the point (5.45,0) $\mu$m. There are two
minima with energies about -8.5 eV at the tips of the maghemite
chain. These minima are the global energy minima, being the spots of
energetically most favorable attachment of the MC to the chain of
MPs. This fact is in agreement with experimental observations, where
the MC was observed at the tip of the chain \cite{Fleissner07a}.

\begin{figure}
\begin{center}
\includegraphics[scale=0.8,clip]{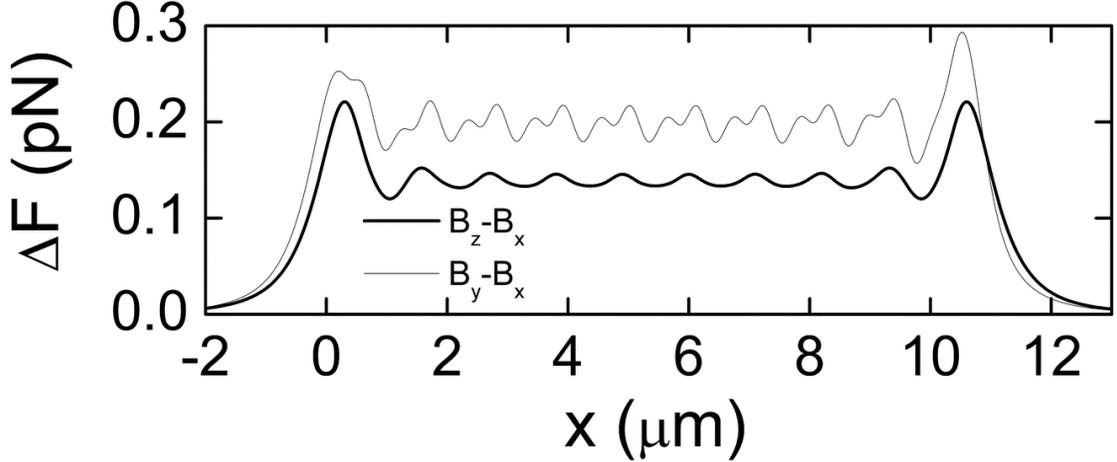}
\end{center}
\caption{Difference in force acting on the MC and arising due to the
90$^{\circ}$ change of the direction of the external magnetic field.
The thin line shows the force difference arising due to the change
of external magnetic field direction from x to z, and thick line
shows the force difference arising due to the change of external
magnetic field direction from x to y. The force differences were
calculated as a function of the x-coordinate of the MC, while y=0.8
$\mu$m and z=0 $\mu$m.} \label{fg:Forces_point_dif}
\end{figure}

To illustrate the effect of the external magnetic field on the
magnetoreceptor system we have calculated the differences in forces
acting on the MC due to the 90$^{\circ}$ change of the direction of
the external magnetic field. The force differences are shown in
Fig.~\ref{fg:Forces_point_dif}. The thin line shows the force
difference arising due to the change of external magnetic field
direction from x to z, and the thick line shows the force difference
arising due to the change of the external magnetic field direction
from x to y. The force differences were calculated as a function of
x-coordinate of the MC, while y=0.8 $\mu$m and z=0 $\mu$m.
Fig.~\ref{fg:Forces_point_dif} shows that the force change, caused
by the 90$^{\circ}$ change of the direction of the external field is
0.1-0.2 pN in both cases.

It was experimentally demonstrated \cite{Fleissner03,Fleissner07a}
that the MCs are connected to the nerve cell membrane. Depending on
the magnetic field strength the magnetite cluster exerts forces on
the membrane and activates mechanosensitive ion channels increasing
the flux of ions into the cell. The ions change the membrane
potential. If the potential is reduced to the threshold voltage
\cite{Lehninger_Biochemistry}, an action potential is generated in
the cell, which opens up hundreds of voltage-gated ion channels in
the membrane. During the millisecond that the channels remain open,
thousands of ions rush into the cell \cite{Lehninger_Biochemistry},
producing a nerve signal to the brain. The mechanosensitive ion
channels influence the time needed for the membrane potential to
reach the threshold value, and thus influence the birds behavior.

A typical example of a mechanosensitive ion channel is the
transduction channel of a hair cell (for review see
Refs.~\cite{Hudspeth00,Markin95,Hamill01}). The opening/closing of
the mechanosensitive ion channel is regulated by the so-called gate,
which is a large biological complex (protein or complex of proteins)
at the edge of the ion channel
\cite{Hudspeth00,Markin95,Hamill01,Corey94}. The gate is connected
to an elastic element, the gating spring
\cite{Hudspeth00,Markin95,Hamill01,Corey94}, transmitting the force
to the gate.

The ion channel has two conformations: closed and open. Because the
gate swings through a distance $\lambda$ upon opening, an external
force $f$ changes the energy difference between open and closed
states and can bias the channel to spend more time in its open
state. The gating springs are connected to the magnetite cluster
which produces an external pull on the gates. As follows from our
calculations the magnitude of this pull is about 0.2 pN, when the
direction of the external magnetic field is changed on 90$^{\circ}$.
The work done in gating the channel is \cite{Corey94}: $ \Delta
E=\Delta\varepsilon -f\lambda$, where the first term represents the
change of the intrinsic energy between the open and the closed
states of the channel and the second term shows the work of external
force required for opening the channel. $\lambda$ is the
displacement of the gate. For the mechanosensitive ion channels in
hair cells $\lambda\approx 4$ nm \cite{Corey94,Hudspeth00}. The
probability for the ion channel to be open in the presence of
external force is:

\begin{equation}
p=\frac{1}{1+\exp\left(\frac{\Delta\varepsilon-f\lambda}{kT}\right)}.
\label{eq:p}
\end{equation}

\noindent If no external force is applied then $f=0$ and the
corresponding probability for the channel to be open is ${\tilde
p_0}$. Thus, the change of channel opening probability due to the
applied force is:

\begin{equation}
\eta=\frac{p-{\tilde p_0}}{{\tilde p_0}}=
\frac{\exp\left(\frac{\Delta\varepsilon}{kT}\right)\left(\exp\left(\frac{f\lambda}{kT}\right)-1\right)}{\exp\left(\frac{f\lambda}{kT}\right)+\exp\left(\frac{\Delta\varepsilon}{kT}\right)}.
\label{eq:eta}
\end{equation}

\noindent The value of $\Delta\varepsilon$ is not known. Usually
\cite{Corey94}, it is assumed that $\Delta\varepsilon=0$, but in
general it is not because the gate can form hydrogen bonds with the
membrane, which break when the gate is opened. Thus
$\Delta\varepsilon>0$.

\begin{figure}
\begin{center}
\includegraphics[scale=0.8,clip]{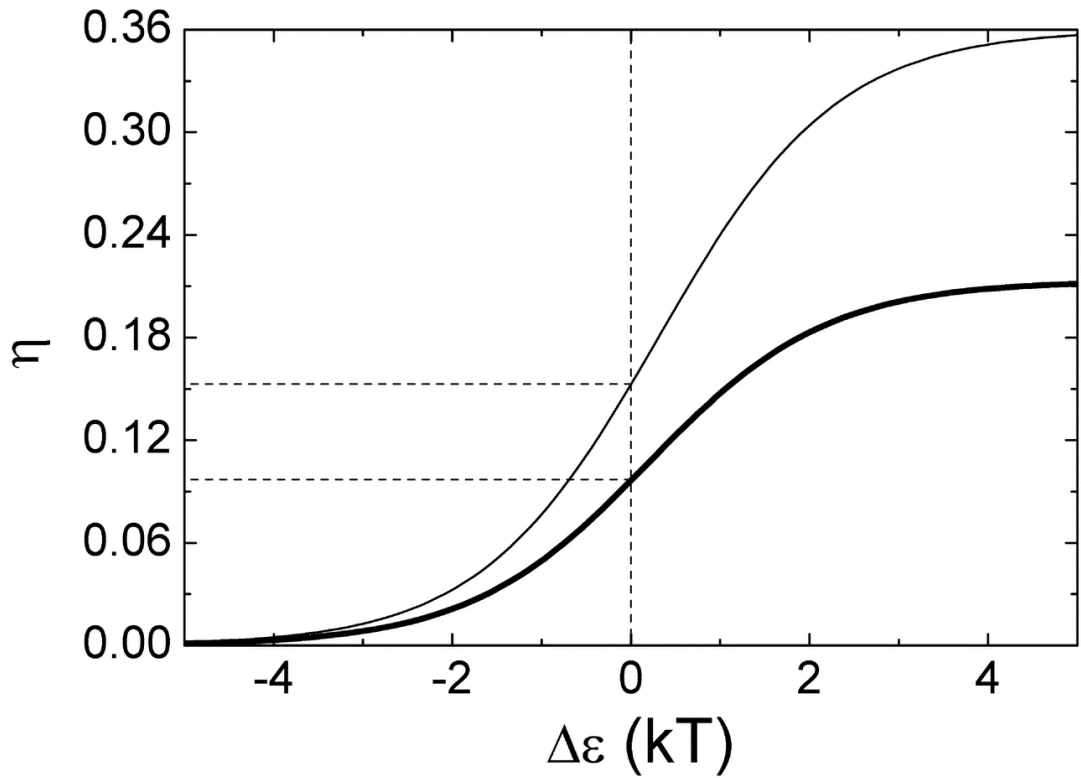}
\end{center}
\caption{Change of the mechanosensitive ion channel opening
probability calculated as the function of change of the intrinsic
energy between the open and the closed states of the channel. The
thick line corresponds to the gating-spring transducer mechanism and
thin line corresponds to the mechanism based on the elastic
deformation of the membrane.} \label{fg:probability}
\end{figure}

Fig.~\ref{fg:probability} shows the dependence of the change of
channel opening probability, $\eta$ on $\Delta\varepsilon$ (thick
line). From Fig.~\ref{fg:probability} and from Eq.~(\ref{eq:eta}) it
follows that the change of channel opening probability saturates at
large values of $\Delta\varepsilon$. The limiting value is
$\eta_{max}=\exp\left(\frac{f\lambda}{kT}\right)-1$. For the given
$f$, $\lambda$ and $T$: $\eta_{max}=0.21$, being the maximal change
of channel opening probability possible in the suggested mechanism.
If $\Delta\varepsilon=0$ then $\eta_0=0.096$. If $\Delta\varepsilon$
is positive then $\eta$ is somewhere between $\eta_0$ and
$\eta_{max}$.

\noindent

Another possible transducer mechanism of the geomagnetic field is
based on the elastic deformation of the membrane. The deformation
mechanism might arise in addition to the gating mechanism or be an
alternative to it. The work performed on membrane deformation is
given by \cite{IliaGreiner07}:

\begin{equation}
A=\gamma \Delta S=\frac{f^2}{\pi\gamma}. \label{eq:workDeform}
\end{equation}

\noindent where $\Delta S$ is the change of the membrane surface
area and $\gamma$ is the membrane surface tension coefficient.
Substituting $A$ instead of $f\lambda$ in Eq.~(\ref{eq:eta}) one
obtains the change in channel opening probability caused by the
membrane deformation.

Fig.~\ref{fg:probability} shows the dependence of the change of
channel opening probability caused by the membrane deformation, on
$\Delta\varepsilon$ (thin line) obtained for $f=0.2$ pN and
$\gamma=0.01$ dyn/cm=$10^{-5}$ N/m, being the typical surface
tension coefficient of a membrane \cite{Hochmuth96,Dai98}. The
maximal value of $\eta^{def}$ is $\eta^{def}_{max}=0.36$, being 1.7
times greater than in the case of the gate-spring mechanism
discussed above. If $\Delta\varepsilon=0$ then $\eta^{def}_0=0.15$.
Since $\Delta\varepsilon$ is expected to be positive then
$0.15<\eta^{def}<0.36$.

In the present paper a possible mechanism of avian orientation in a
magnetic field is discussed. It was shown that in the external
magnetic field the MCs experience an attractive (repulsive) force
leading to their displacement, which induces a primary receptor
potential via mechanosensitive membrane channels leading to a
certain orientation effect of a bird. We believe that the suggested
magnetoreception mechanism is a realistic candidate for the
magnetoreception mechanism in birds. It might also be responsible
for magnetosensation in other animals like fishes, salamanders, bees
(for review see Ref.~\cite{IliaGreiner07}). Unfortunately, lack of
sufficient information about magnetic particles in these species
hinders us to draw conclusions about their precise magnetoreception
mechanism. However, we believe, that the magnetoreception mechanism
should be universal, i.e. the same for all kinds of animals with,
probably, minor alternations. Therefore, after more experimental
data regarding the magnetic particles in animals become available
the present investigation can be easily extended to a more general
description.

We thank Professors Gerta and G{\"u}nther Fleissner for many helpful
discussions. We also thank Professors Klaus Schulten and Andrey
Solov'yov, as well as Mr. Alexander Yakubovich for many insightful
comments. We acknowledge support of this work by the NoE EXCELL.

\end{document}